\documentclass{emulateapj}

\def\mdot{$\dot M$}

\def\xte{{\it RXTE~}} 

\shortauthors{Lin et al.}

\begin{document}

\title{Evaluating Spectral Models and the X-ray States of Neutron-Star X-ray Transients}

\author{Dacheng Lin\altaffilmark{1}, Ronald A. Remillard, and Jeroen Homan}
\affil{MIT Kavli Institute for Astrophysics and Space Research, MIT, 70 Vassar Street, Cambridge, MA 02139-4307}

\altaffiltext{1}{email: lindc@mit.edu}

\begin{abstract}
 
We analyze the X-ray spectra of the neutron-star (NS) X-ray transients
\object{Aql X-1} and \object{4U 1608-52}, obtained with {\it RXTE}
during more than twenty outbursts. Our aim is to properly decompose
the spectral components and to study their evolution across the hard
and soft X-ray states. We test commonly used spectral models and
evaluate their performance against desirability criteria, including
$L_{\rm X} \propto T^4$ evolution for the multicolor disk (MCD)
component, and similarity to black holes (BHs) for correlated
timing/spectral behavior. None of the classical models for thermal
emission plus Comptonization perform well in the soft state. Instead,
we devise a hybrid model: for the hard state a single-temperature
blackbody (BB) plus a broken power-law (BPL) and for the soft state
two thermal components (MCD and BB) plus a constrained BPL. This model
produces $L_{\rm X} \propto T^4$ tracks for both the MCD and BB, and
it aligns the spectral/timing correlations of these NSs with the
properties of accreting BHs.  The visible BB emission area is very
small ($\sim$ $1/16$ of the NS surface), but it remains roughly
constant over a wide range of $L_{\rm X}$ that spans both the hard and
soft states. We discuss implications of a small and constant boundary
layer in terms of the presence of an innermost stable circular orbit
that lies outside the NS. Finally, if the BB luminosity tracks the
overall accretion rate, then we find that the Comptonization in the
hard state has surprisingly high radiative efficiency, compared to MCD
emission in the soft state.  Alternatively, if we assume that the
radiative efficiency of a jet in the hard state must be less than the
MCD efficiency in the soft state, while relaxing presumptions about
the accretion rate, then our results may suggest substantial mass
outflow in the jet.

\end{abstract}


\section{INTRODUCTION}

In low-mass X-ray binaries (LMXBs), a neutron star (NS) or
stellar-mass black hole (BH) accretes matter from a Roche-lobe
filling, low-mass companion star through an accretion disk. X-rays
are produced by the inner accretion disk and/or the boundary layer
formed by impact of the accretion flow with the NS surface. The
luminous and weakly magnetized NS LMXBs are classified into atoll and
Z sources based on their X-ray spectral and timing properties
\citep{hava}. In a color-color diagram, Z sources trace out roughly
Z-shaped tracks within hours to a day or so. Their X-ray spectra are
normally soft in all three branches of the ``Z'', i.e., most of the
flux is emitted below 20 keV.  Atoll sources, however, show more
dramatic spectral changes, albeit on longer time scales (days to
weeks); their spectra are usually soft at high luminosities and hard
when they are faint.

Recently, it was found that some atoll sources can also exhibit
Z-shaped tracks in the color-color diagram when they are observed over
a large range of luminosity \citep{murech2002, gido2002a}.  However,
it was noted by several authors \citep{baol2002,
vavame2003,revava2004,va2006} that the properties of atoll sources
(e.g., rapid X-ray variability, the order in which branches are traced
out) are very different from those of the Z sources. The spectral
states of atoll sources in the upper, diagonal and lower branches of
these Z-shaped tracks are often referred to as the ``extreme island'',
``island'', and ``banana'' states/branches,
respectively. However, in this paper we will use the terms ``hard'',
``transitional'', and ``soft'' states, respectively.

The spectral modeling of accreting NSs has been controversial for a
long time \citep[see][for a review]{ba2001}. In the soft state, the
spectra are generally described by models that include a soft/thermal
and a hard/Comptonized component. Based on the choice of the thermal
and Comptonized components, there are two classical models, often
referred to as the {\it Eastern} model \citep[after][]{miinna1989}
and the {\it Western} model \citep[after][]{whstpa1988}. In the {\it
Eastern} model, the thermal and Comptonized components are described
by a multicolor disk blackbody (MCD) and a weakly Comptonized blackbody,
respectively. In the {\it Western} model, the thermal component is a
single-temperature blackbody (BB) from the boundary layer and there
is Comptonized emission from the disk. The color temperature for the
thermal component (i.e., the temperature at the inner disk radius
$kT_{\rm mcd}$ for a disk model or $kT_{\rm bb}$ for a boundary layer
model) is typically in the range of $\sim$ 0.5--2.0 keV
\citep[e.g.,][]{baolbo2000,oobagu2001, diiabu2000, iadiro2005}. For
the Comptonized component, the same authors reported a plasma
temperature of $\sim$ 2--3 keV and a large optical depth of $\sim$
5--15 in the soft state (for a spherical geometry).

In the hard state, the spectra are dominated by a hard/Comptonized
component, but a soft/thermal component is generally still required
\citep{chsw1997, baolbo2000, chba2001,gido2002b}. The thermal
component can be either a BB or a MCD, but the latter seems to be
ruled out by the inferred inner disk radii that are unphysically small.
The color temperature of this component is typically $\lesssim$ 1 keV
\citep[e.g.,][]{baoloo2003,chba2001}. The inferred plasma electron
temperature of the Comptonized component is typically a few tens of
keV and its optical depth is $\sim$ 2--3 in the hard state (for a
spherical geometry).  The hard state of atoll sources is considered
by some authors to be associated with a steady jet
\citep[e.g.,][]{fe2006,mife2006}. The inverse Compton spectrum could
then arise from the base of the jet, and synchrotron emission would
contribute seed photons and perhaps a secondary contribution to the
X-ray spectrum \citep{manowi2005}.

In the past, various approaches have been made to further our
understanding of X-ray spectra of accreting NSs. These include: (1)
spectral surveys of a large number of sources, covering a wide range of
luminosities \citep{chba2001,chsw1997}, (2) detailed studies of a
large number of observations from a single outburst of a NS X-ray transient
\citep{gido2002b,maco2003b,maba2004}, (3) Fourier frequency resolved
X-ray spectroscopy \citep{giremo2003,olbagi2003}, and (4) comparisons
of spectral and timing properties to those of BH LMXBs, to understand
which features might be the result of the presence/absence of a solid
surface \citep{wi2001,ba2001,dogi2003}. However, a general consensus
on the appropriate X-ray spectral model for the various subtypes and
states of accreting NSs has not been achieved.

In this paper we present an extensive study of a large number of
observations of two NS transients \object{Aql X-1} and \object{4U
1608-52}, using data obtained with the {\it Rossi X-ray Timing
Explorer} \citep[\xte,][]{brrosw1993}, during more than twenty
individual outbursts. There are several advantages of using this
archive. First, it allows us to compare the evolution of spectral
properties of different outbursts systematically. Second, we can
examine the behavior of a specific spectral component with changes in
accretion rate, capitalizing on the large range in luminosity
exhibited by these atoll-type transients. Third, compared with surveys
using just a few observations but many sources, we can reduce the
problems due to our poor knowledge of the parameters of the sources
(e.g., the distance, inclination, and absorption).

The goal of this paper is to find a spectral model that can well
describe the X-ray spectra of NS X-ray transients. Although the NS
X-ray transients often change luminosity by several orders of
magnitude and show substantial diversity in outbursts, their well
organized color-color and color-intensity diagrams clearly show that
their spectral evolution tracks are narrow and thus repeatable. This
compels the efforts to unlock spectral models for NS systems to
explain the well-behaved sources in terms of accretion physics, and
also to further investigate the differences and similarities between
BHs and NSs.

\object{Aql X-1} and \object{4U 1608-52} have been classified as
atoll sources \citep{hava, remeva2000}. Recently, \citet{revava2004}
and \citet{vavame2003} analyzed the general timing properties of
these two sources. Here we will concentrate on their spectral
properties. We describe our data reduction scheme in
\S\ref{sec:reduction} and show the long-term light curves and
color-color diagrams in \S\ref{sec:colordiag}. We perform and
evaluate detailed spectral modeling in \S\ref{sec:specmod}. In
\S\ref{sec:combh}, we compare timing properties with BHs in order to
further evaluate the spectral models. We argue that a particular
model is most suitable for these NS transients, and then we further
explore the ramifications of this model for X-ray states and the
physical properties of accretions in \S\ref{sec:propertys} and
\S\ref{sec:propertyh}. Finally we give our summary and discussion.

\section{OBSERVATIONS AND DATA REDUCTION}
\label{sec:reduction}

For our analysis, we used all of the available \xte observations of
\object{Aql X-1} and \object{4U 1608-52} prior to 2006 January 1.
Data were analyzed from the Proportional Counter Array
\citep[PCA;][]{jaswgi1996} and the High Energy X-ray Timing Experiment
\citep[HEXTE;][]{roblgr1998} instruments. We utilized the
best-calibrated detector units of each instrument, which are
Proportional Counter Unit 2 (PCU 2) for the PCA and Cluster A for the
HEXTE, and we extracted the average pulse-height spectra, one for each
\xte observation. All spectral extractions and analyses utilized the
FTOOLS software package version 6.0.4. Some standard criteria were
used to filter the data: data of 20 seconds before and 200 seconds
after type I X-ray bursts were excluded \citep[see][]{relico2006}; the
earth-limb elevation angle was required to be larger than $10\degr$;
the spacecraft pointing offset was required to be $< 0.02\degr$. For
faint observations, we additionally excluded data within 30 minutes of
the peak of South Atlantic Anomaly passage or with large trapped
electron contamination.

We only considered observations that yielded data from the PCA, HEXTE
and relevant spacecraft telemetry. Only observations with PCA
intensity (background subtracted) larger than 10 counts/s/PCU were
used.  We required the exposure of the spectra to be larger than five
minutes for faint observations (source intensity lower than 40
counts/s/PCU) and two minutes for bright observations. Appropriate
faint/bright background models were used when the source had intensity
lower or higher than 40 counts$/$s$/$PCU. For the PCA, the spectra
were extracted from ``standard 2'' data collection mode and the
response files were created so that they were never offset from the
time of each observation by more than 20 days. For the HEXTE, the
program HXTLCURV was used for spectral extraction, background
subtraction, and deadtime correction. Finally, we applied systematic
errors of $0.8\%$ for PCA channel 0--39 (about below 18 keV) and $2\%$
for PCA channel 40--128 \citep{krwico2004,jamara2006}. No systematic
errors were applied for HEXTE data. A summary of the observations and
other source properties is given in Table~\ref{tbl-1}. All further
analyses and spectral fits were uniformly applied to each selected
observation.

\section{LIGHT CURVES AND COLOR-COLOR DIAGRAMS}
\label{sec:colordiag}

\begin{figure*} \epsscale{1.0}  
%\figurenum{1}  
\plotone{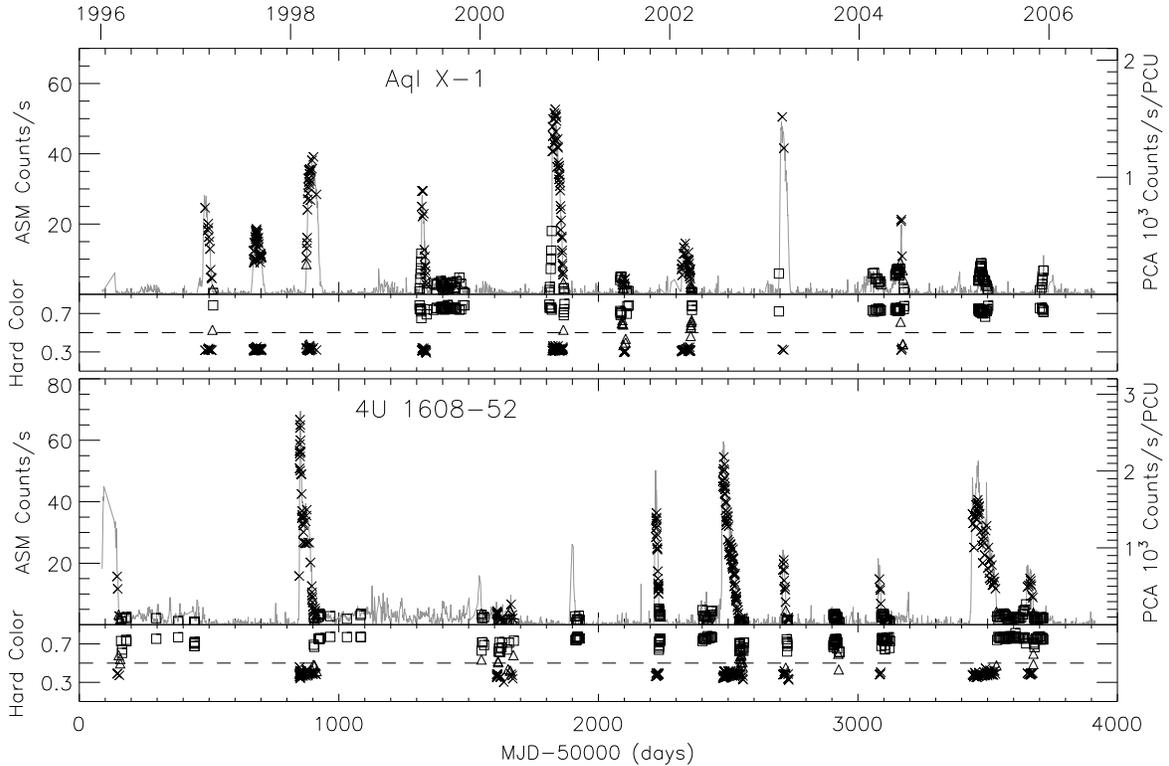}
\caption{Long-term light and color curves of Aql~X--1 and 4U~1608--52,
showing a large variety of outburst properties.  Grey solid lines are
from \xte ASM and discrete symbols are from PCA observations representing
different spectral states: hard (square), transitional (triangle), and
soft (cross). The dashed line (hard color = 0.5) is a reference
to help distinguish the different states.
\label{fig:asmdata}} 
\end{figure*}

\begin{figure} 
%\figurenum{2} 
\plotone{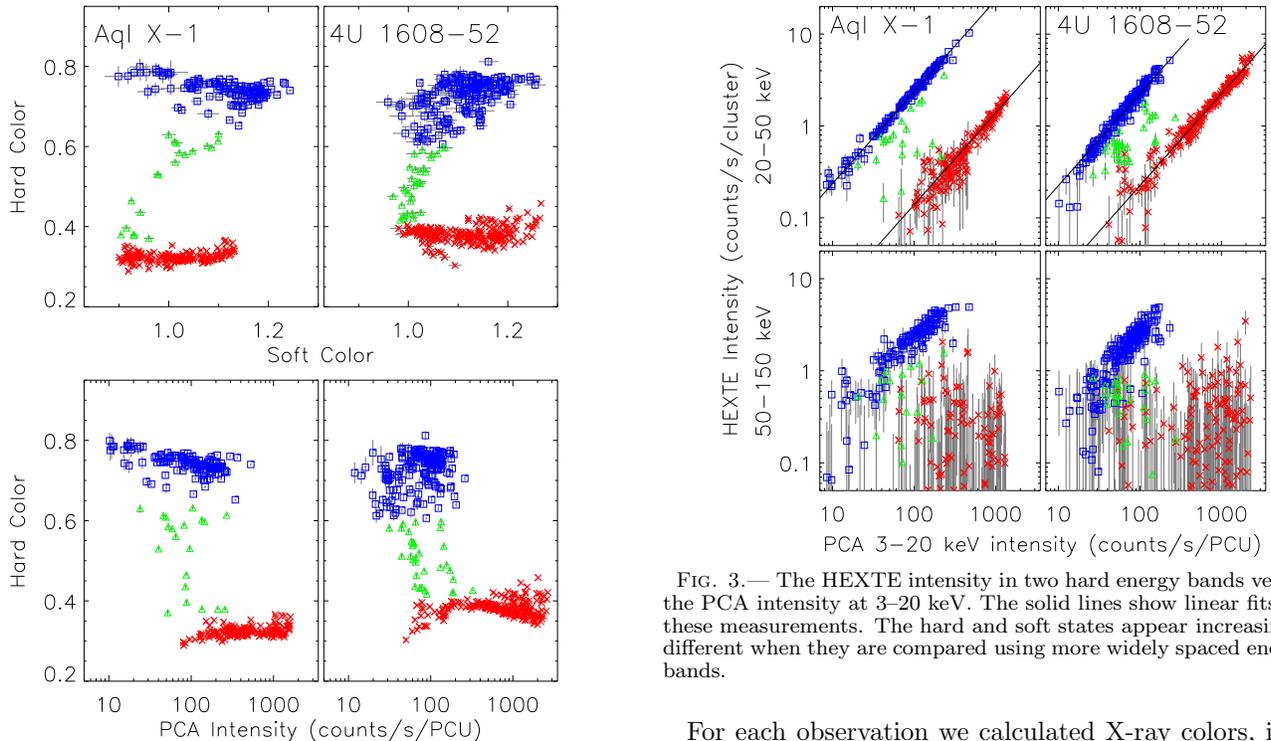} 
\caption{Normalized color-color and color-intensity diagrams of
Aql~X--1 and 4U~1608--52 from ten years of pointed \xte PCA
observations. The PCA intensity is the normalized count rate from
PCU2, and in these units the Crab Nebula yields 2520 c/s.  The hard,
transitional, and soft states are represented by blue squares, green
triangles, and red crosses, respectively. The uncertainties are at 1
$\sigma$ confidence and are normally smaller than the symbol size.
\label{fig:ccdiag}}
\end{figure}

\begin{figure}
%\figurenum{3}
\plotone{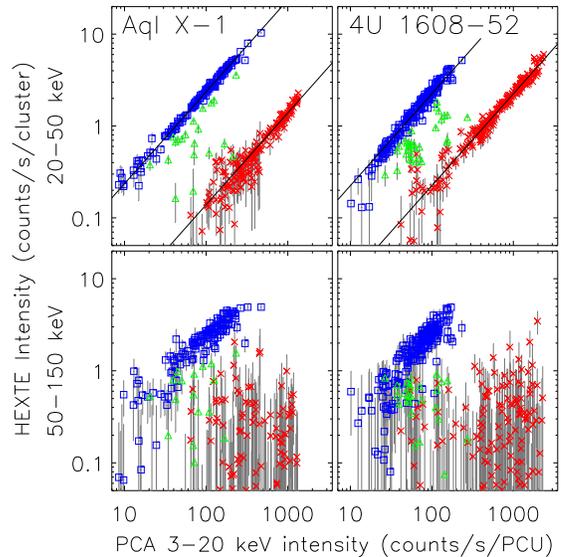} 
\caption{The HEXTE intensity in two hard energy bands versus the PCA intensity
at 3--20 keV. The solid lines show linear fits for these measurements. The
hard and soft states appear increasingly different when they are
compared using more widely spaced energy bands.
\label{fig:pcahexcr}}
\end{figure}

The long-term light and color curves of \object{Aql X-1} and
\object{4U 1608-52} are shown in Figure~\ref{fig:asmdata}. This
figure combines data from the PCA and the \xte All-Sky Monitor
\citep[ASM;][]{lebrcu1996}. Typically, outbursts start in the hard
state, evolve to the soft state, and finally return to the hard state
during the decay. However, outbursts can be quite different from each
other, e.g., in amplitude and duration. Moreover, in some outbursts
the source did not enter the soft state.

For each observation we calculated X-ray colors, in a manner similar
to \citet{murech2002}. Soft and hard colors were defined as the
ratios of the background-subtracted counts in the
(3.6--5.0)/(2.2--3.6) keV bands and the (8.6--18.0)/(5.0--8.6) keV
bands, respectively. We normalized the raw count rates from each
PCU with the help of observations of the Crab Nebula.  For each PCA
gain epoch, we computed linear fits (vs. time) to normalize the Crab
count rates to target values of 550, 550, 850, and 570 counts/s/PCU in
these four energy bands.  The normalized color-color and color-intensity
diagrams in Figure~\ref{fig:ccdiag} resemble those in
\citet{murech2002}, although our figure includes more data and we used
a different normalization scheme for the softest energy bands.
Although these two sources are atoll sources, their tracks in the
color-color diagrams are Z-shaped, owing to their large range of
luminosities. We combined the hard color and PCA intensity to
pragmatically define the source states as shown in
Figure~\ref{fig:ccdiag}. The hard state has a hard color $> 0.65$
(\object{Aql X-1}) or $> 0.6$ (\object{4U 1608-52}); the soft state
has a PCA intensity $> 400$ counts/s/PCU and a hard color $< 0.5$ or a
PCA intensity $< 400$ counts/s/PCU and a hard color $< 0.36$
(\object{Aql X-1}) or $< 0.41$ (\object{4U 1608-52}). All of the
remaining observations are referred to as the transitional state. For
all figures in this paper, the hard, transitional, and soft states are
represented by blue squares, green triangles, and red crosses,
respectively. We note that strong hysteresis is observed in both
sources, that is, the hard-soft transition generally occurs at higher
X-ray flux compared to the soft-hard transition
\citep[e.g.,][]{maco2003a}. The transitional-state observations are
mostly from the decay phases of the outbursts, since the rise is often
sparsely covered.

Figure~\ref{fig:pcahexcr} shows the relation of source intensities in
more widely spaced energy bands. The top panels are the HEXTE 20--50 keV
intensity versus the PCA 3--20 keV intensity. The most striking aspect of
these panels is that there are two nearly linear tracks corresponding
to the soft and hard states.  The bottom panels are the HEXTE 50--150 keV
intensity versus the PCA 3--20 keV intensity. We can still see the linear
track in the hard state, but the sources are generally not detected
above 50 keV in the soft state.

While the light curves in Figure~\ref{fig:asmdata} show substantial
diversity in outburst amplitude and duration for a given source,
Figures~\ref{fig:ccdiag} and \ref{fig:pcahexcr} show that the
superposition of all observations on color-color and color-intensity
diagrams shows well organized spectral states and very common
behaviors in these two sources. The aim of this paper is to
capitalize on this organization and the comprehensive \xte data
archive, in order to determine the best way to model the X-ray spectra
across these different states.

\section{SPECTRAL MODELING}
\label{sec:specmod}

\subsection{Spectral Models and Assumptions}

In this work, we fitted the X-ray spectra of \object{Aql X-1} and
\object{4U 1608-52} with several different models. The PCA and HEXTE
pulse-height spectra were fitted jointly over the energy range
2.6--23.0 keV and 20.0--150.0 keV, respectively, allowing the
normalization of the HEXTE spectrum, relative to the PCA spectrum, to
float between 0.7 and 1.3. For soft-state observations, HEXTE spectra
were used up to 50 keV because the flux at higher photon energies was
negligible (Figure~\ref{fig:pcahexcr}).

In the typical description of the NS continuum spectra, some form
of pure thermal radiation is often combined with another radiation
process, commonly presumed to be some form of Comptonization, although synchrotron
radiation might also be involved (\S 1). Hereafter, we simply
refer to emission other than the pure thermal radiation as the
``Comptonized'' component. Two forms of thermal radiation were
considered: BB and MCD models (bbodyrad and diskbb in XSPEC
respectively). The BB model provides the color temperature ($kT_{\rm
bb}$) and the apparent radius ($R_{\rm bb}$; isotropic assumption) of
the BB emission area, while the MCD model provides the apparent inner
disk radius ($R_{\rm mcd}$) and the color temperature at the inner
disk radius ($kT_{\rm mcd}$).

As for the modeling of the Comptonized component, we considered
both a broken power-law model (bknpower in XSPEC, hereafter BPL) and
the Comptonization model by \citet{ti1994} (CompTT in XSPEC). The
CompTT model computes the Comptonization of a Wien input spectrum of
"seed photons" by a hot (single temperature) plasma with a uniform
covering geometry.  On the other hand, the BPL component can be
considered as a functional approximation for Comptonization under
complex conditions or in combination with another radiation process
like synchrotron radiation. We gave these two models an equal
opportunity to handle the effects of Comptonization when we tested
different kinds of spectral decomposition. Thus BPL or CompTT was
combined with BB and/or MCD in a variety of ways, as summarized in
Table~\ref{tbl-2}. The BPL model has four parameters: two photon
indices, a break energy ($E_{\rm b}$) and a normalization
parameter. The CompTT model is parametrized by the seed-photon
temperature ($kT_{\rm s}$), the plasma electron temperature ($kT_{\rm
e}$), the optical depth ($\tau$), a parameter describing the geometry
of the Comptonizing cloud (either spherical or disk-like), and a
normalization parameter.

 All models also included a Gaussian line. Its central line energy was
constrained to be between 6.2--7.3 keV, targeting the Fe line 6.4 keV
\citep{asdona2000}. The average best-fitting value was $\sim$ 6.6 keV
for both sources. The intrinsic width of the Gaussian line ($\sigma$)
was fixed at $0.1$ keV. This is consistent with the {\it ASCA} result
of \citet{chba2001}. The PCA has energy resolution $\sim$ 1 keV,
limiting the need for a precise value. An interstellar absorption
component was also included with the hydrogen column fixed at $N_{\rm
H}=0.5\times 10^{22}$ ${\rm cm}^{-2}$ for \object{Aql X-1}
\citep{chba2001} and $1.0\times 10^{22}$ ${\rm cm}^{-2}$ for
\object{4U 1608-52} \citep{pedava1989}. Since no eclipses or
absorption dips have been observed, the two sources are likely not
high-inclination systems and we assumed their binary inclinations to
be $60\degr$. We scaled the luminosity and radius related quantities
using distances of 5 kpc for \object{Aql X-1} \citep{rubibr2001} and
3.6 kpc for \object{4U 1608-52} \citep{nabaco2000}, unless indicated
otherwise.

\subsection{Comptonized + Thermal Two-component Models}

\subsubsection{The Problem of Model Degeneracy}
\label{moddeg}

\begin{figure}
%\figurenum{4} 
\plotone{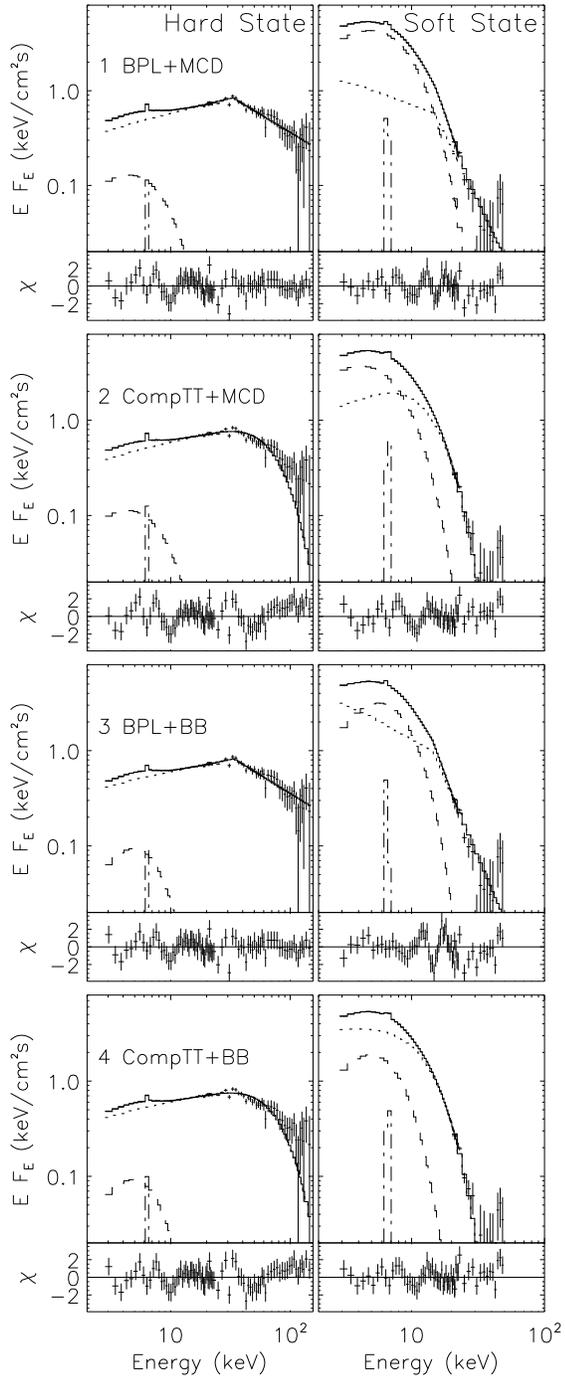}
\caption{The unfolded spectra and residuals of two sample observations
of Aql~X--1 using different kinds of models. One is in the hard state
and the other in the soft state. The panels for models with CompTT are
from spectral fits with seed-photon temperatures $\lesssim 0.5$ keV
(cold-seed-photon models). The total model fit is shown as a solid
line, the Comptonized component as a dotted line, the thermal
component as a dashed line, and the Gaussian line as a dot-dashed
line.
\label{fig:samspec}}
\end{figure}
\begin{figure}
%\figurenum{5} 
\plotone{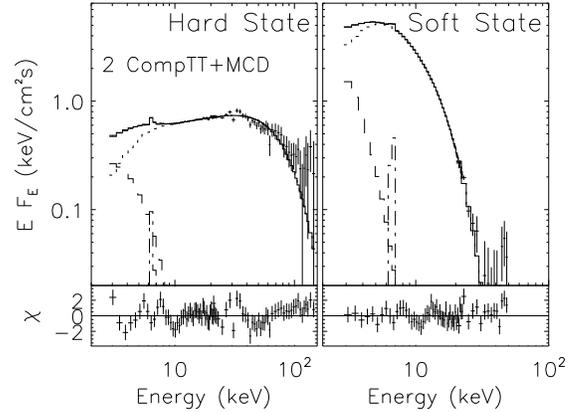}
\caption{The unfolded spectra and residuals of two sample observations
 of Aql~X--1 with best-fitting $kT_{\rm s}$ $\gtrsim$ 1 keV using
 Model CompTT+MCD (hot-seed-photon model). The total model fit is
 shown as a solid line, the Comptonized component as a dotted line,
 the thermal component as a dashed line, and the Gaussian line as a
 dot-dashed line.
\label{fig:samspechot}}
\end{figure}

First we consider models with two continuum components: one is
Comptonized and the other is thermal (Models 1--4 in
Table~\ref{tbl-2}). It turns out that the spectral fitting is
inherently non-unique. To illustrate the contribution of each
component and the problem of model degeneracy, we use two observations
of \object{Aql X-1}.  One is a hard-state observation on 2004 February
21 (hard color 0.73, intensity 178.5 counts/s/PCU), the other is a
soft-state observation on 2000 October 27 (hard color 0.33, intensity
1262 counts/s/PCU).

Figures~\ref{fig:samspec} and \ref{fig:samspechot} show the unfolded
spectra of these two observations using different Comptonized +
thermal models. All models give acceptable spectral fits, despite the
fact that the spectra each contain well over $10^6$ counts.
Figure~\ref{fig:samspec} shows that there is not only degeneracy from
the choice of the thermal component (i.e., BB or MCD), but also from
the choice of the Comptonized component (i.e., BPL or CompTT).
Moreover, even if we choose Model CompTT+MCD or CompTT+BB, there are
still two competing $\chi^2$ minima, one with best-fitting $kT_{\rm
s}$ $\lesssim 0.5$ keV and the other with best-fitting $kT_{\rm s}$
$\gtrsim 1$ keV. Their corresponding unfolded spectra can be seen in
Figures~\ref{fig:samspec} and \ref{fig:samspechot}, respectively.
Table~\ref{tbl-3} gives the detailed results using the CompTT+MCD and
CompTT+BB models for our two representative spectra. Hereafter, we
call models with best-fitting $kT_{\rm s}$ $\gtrsim$ 1 keV
``hot-seed-photon models'' and models with best-fitting $kT_{\rm
s}\lesssim 0.5$ keV ``cold-seed-photon models''. Hot-seed-photon
models require that the modeling of the Comptonized component takes
into account the spectral curvature expected from having the seed
photons close to the observed bandpass \citep{dozysm2002}. We note
that the temperature of the chosen thermal component increases and its
normalization decreases significantly when the seed-photon model flips
from the hot to cold solution, for a given observation.

Examination of the other observations shows that this seed-photon
problem is quite general. CompTT+MCD and CompTT+BB models
(Table~\ref{tbl-2}) yield two solutions each: photon temperature
$kT_{\rm s}$ $\gtrsim$ 1 keV and $kT_{\rm s}\lesssim 0.5$ keV.  The
inferred parameters of the thermal component are also quite different
(Table~\ref{tbl-3} and references following). For the soft-state
observations, the inferred temperature of the thermal component is
typically $> 1$ keV for the cold-seed-photon models
\citep[e.g.,][]{baolbo2000, oobagu2001, whstpa1988}.  Otherwise, if
the hot-seed-photon models are used, the temperature of the thermal
component is normally $< 1$ keV and thus less than $ kT_{\rm s}$
\citep[e.g.,][]{diiabu2000, distro2000, iadiro2005}. For the
hard-state observations, if BB is used as the thermal component, the
size of the BB emission area is typically very small, $\sim 2$ km, for
the cold-seed-photon models
\citep[e.g.,][]{baoloo2003,chba2001,gido2002b}.  However, with the
hot-seed-photon models, the BB emission area can be comparable to the
size of the NS \citep[e.g.,][]{baolbo2000, nabaco2000, gupase1998}.
Replacement of BB with MCD in cold-seed-photon models yields very
small inner disk radii as already found by other authors (see
references in \S1). We also point out that many of the references
cited above made use of broad band spectra from {\it BeppoSAX}.  This
means that the seed-photon problem is also present for instruments
that have extended low-energy spectral coverage (see also
\citet{fafrza2005}). We also analyzed the {\it BeppoSAX} observations
of \object{Aql X-1} and \object{4U 1608-52} and found the same problem
to be present.

The low-energy limit of PCA is 2.3--3.0 keV, depending on the gain
setting epoch. We cannot constrain $kT_{\rm s}$ in the
cold-seed-photon models because the peak energy flux of the
seed-photon spectrum (Wien approximation) is at $3 kT_{\rm s} \lesssim
1.5$ keV.  On the other hand, for the hot-seed-photon models, the fits
to PCA spectra produce a cool thermal component in the soft state
making it is difficult to constrain the temperature and normalization
of this component, and sometimes the thermal component is even not
required \citep{gido2002b}.

\subsubsection{Efforts to Resolve Model Degeneracy} \label{resmoddeg}
As outlined in the previous section, there is a degeneracy in X-ray
spectral models for NS LMXBs. The problem is based on the fact that
acceptable fits ($\chi^2_{\nu}$ criteria) can be obtained for either
the hard or soft states by using any combination of two ambiguous
components: a thermal spectrum (BB or MCD) plus a Comptonized
component (BPL, hot-seed CompTT, or cold-seed CompTT). The various
models convey (very) different pictures for the structures and
energetics of NS accretion.

In our descriptions of various model details, we have begun to mention
arguments that have been offered to evaluate competing models in terms
of physical implications derived from the fitted spectral results.
This strategy of performance-based evaluations of spectral models is
most effective when there is a self-consistency issue at stake.
Multiple evaluation criteria will help us to choose a spectral model
that may be superior in overall suitability. Useful considerations can
involve single parameters, spectral evolution, and the relationship
between the hard and soft states. The comparison between the atoll
sources and BHs can provide further constraints on this choice. Below
we list one such consideration used in the literature, followed by
three considerations offered in this paper.  We describe them in term
of problems that are encountered in a particular model/spectral state,
and we track these problems in the fourth and seventh columns of
Table~\ref{tbl-2}.

\begin{enumerate}
\item ``R'' problem: the inner disk radius in models with MCD is too
small, i.e., much less than the size of the NS \citep{chba2001,gido2002b}.
\item ``L'' problem: $L_{\rm mcd} \propto T_{\rm mcd}^4$ (i.e.,
constant inner disk radii) for MCD component is not satisfied for any
meaningful range of luminosity in the soft state; We note that BH
systems do show $L_{\rm mcd} \propto T_{\rm mcd}^4$ when they are in
the soft (thermal) state \citep{kudo2004}.
\item ``T'' problem: $kT_{\rm s} > kT_{\rm bb}$ in models with BB; if
we attribute the BB component to the boundary layer and accept that
the boundary layer should have a higher temperature than the disk
\citep{miinna1989, posu2001}, $kT_{\rm s}$ seems unlikely to be larger
than $kT_{\rm bb}$.
\item ``P'' problem: the Comptonization fraction is not consistent
with the power density spectrum, assuming that a comparison of atoll
sources and BHs in timing properties is relevant. We will explain this
in more detail in \S\ref{sec:combh}.
\end{enumerate}

In the definition of the ``L'' problem, we use the phrase ``over some
range of luminosity'' in recognition of the possibility that the disk
may deviate from the MCD's geometric assumptions, as do BH accretion
disks at high $L_{\rm mcd}$ \citep{kudo2004}.
To help evaluate the ``R'' problem, we inferred the NS radii of our
two sources from the spectral fitting to Type I X-ray bursts (denoted
as $R_{burst}$). During the decay phase of some bursts, an
approximately constant burst emission area is derived when temporal
series of the burst are fitted with a BB model after subtracting the
average persistent emission to isolate the burst from other radiation
components. Such an emission area is expected to be roughly similar to
the apparent size of the NS \citep{levata1993}. Our fits using PCA
data gave $R_{\rm burst}$ $\sim$ 8 km for \object{Aql X-1} and $\sim$
7.2 km for \object{4U 1608-52} (at distances in Table~\ref{tbl-1}),
which are consistent with previous results \citep[e.g.,][]{koinma1981,
nadoin1989}. However, we note that these values are given without any
corrections, e.g., hardening factors and surface occultation by the
inner disk. These issues are further discussed in
\S\ref{sec:propertys}.

\subsubsection{Model Fit Results}

\begin{figure}
%\figurenum{6} 
\plotone{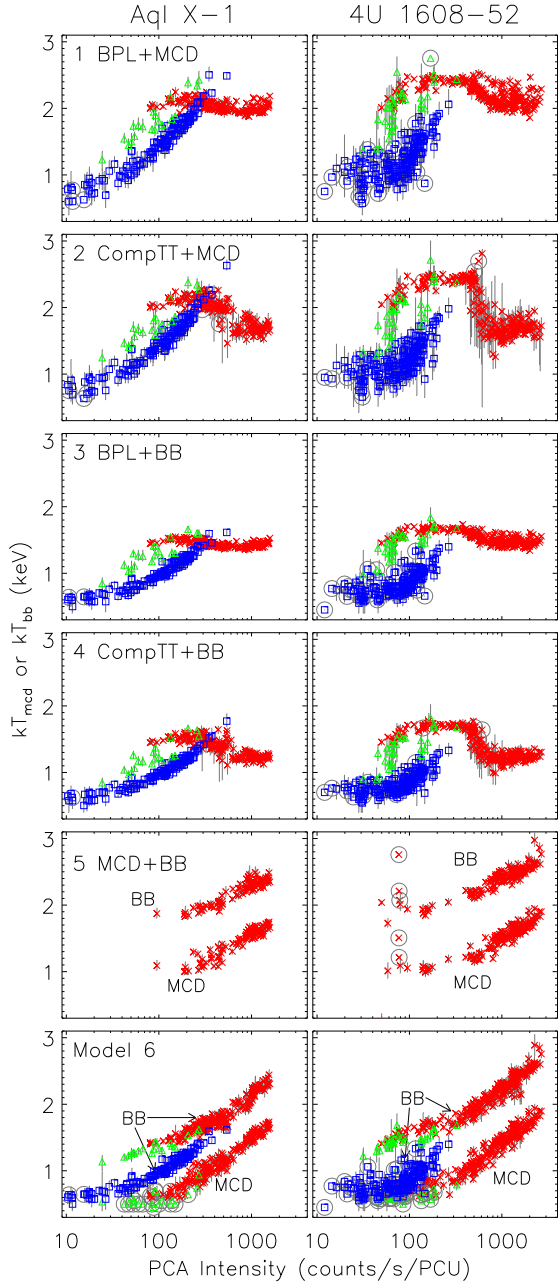}
\caption{Variation of the temperature of the thermal component
($kT_{\rm bb}$ for BB and $kT_{\rm mcd}$ for MCD) with the PCA
intensity. Cold-seed-photon solutions are shown for Models with
CompTT.
\label{fig:tcr}}
\end{figure}

\begin{figure}
%\figurenum{7} 
\plotone{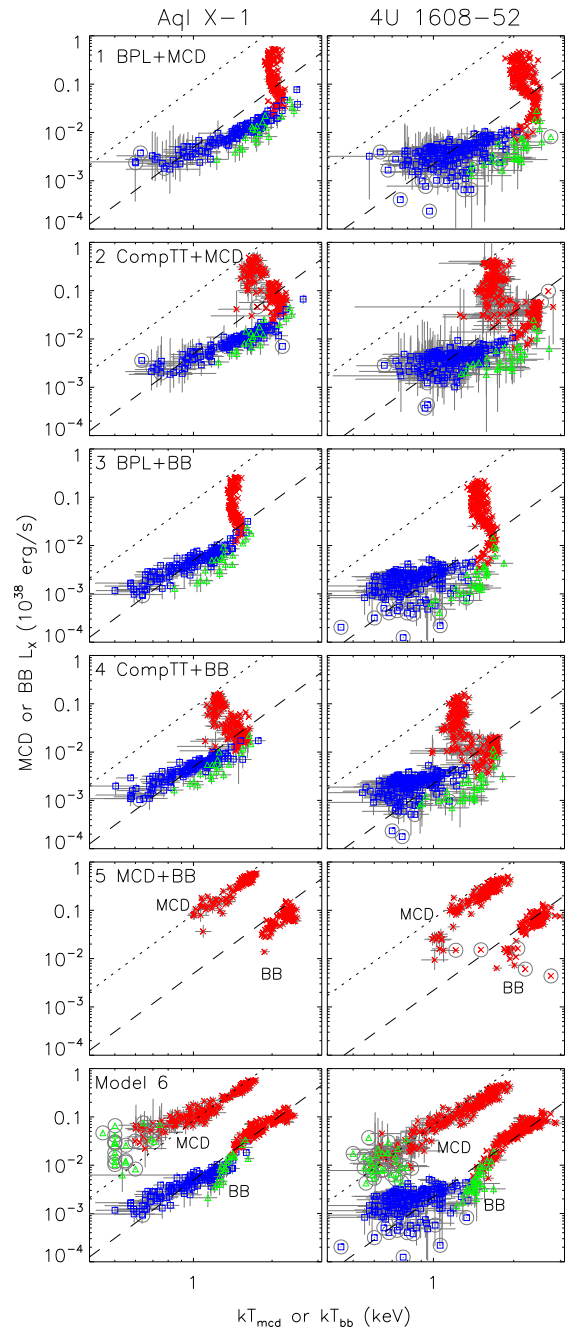}
\caption{The luminosity of the thermal component versus its
characteristic temperature. Models 5 and 6 produce luminosity
evolution for BB and MCD components in the soft state (red crosses)
that are nearly parallel to reference lines, which show $L_{\rm X}
\propto T^4$ with constant emitting surface area. The dotted lines
correspond to the NS burst radii (\S\ref{resmoddeg}) and the dashed lines
correspond to $R =$ 1.9 km and 1.3 km for Aql~X--1 and
4U~1608--52, respectively, assuming $L_{\rm X}=4\pi R^2\sigma
T^4$.
\label{fig:Tlum}}
\end{figure}

\begin{figure}
%\figurenum{8} 
\plotone{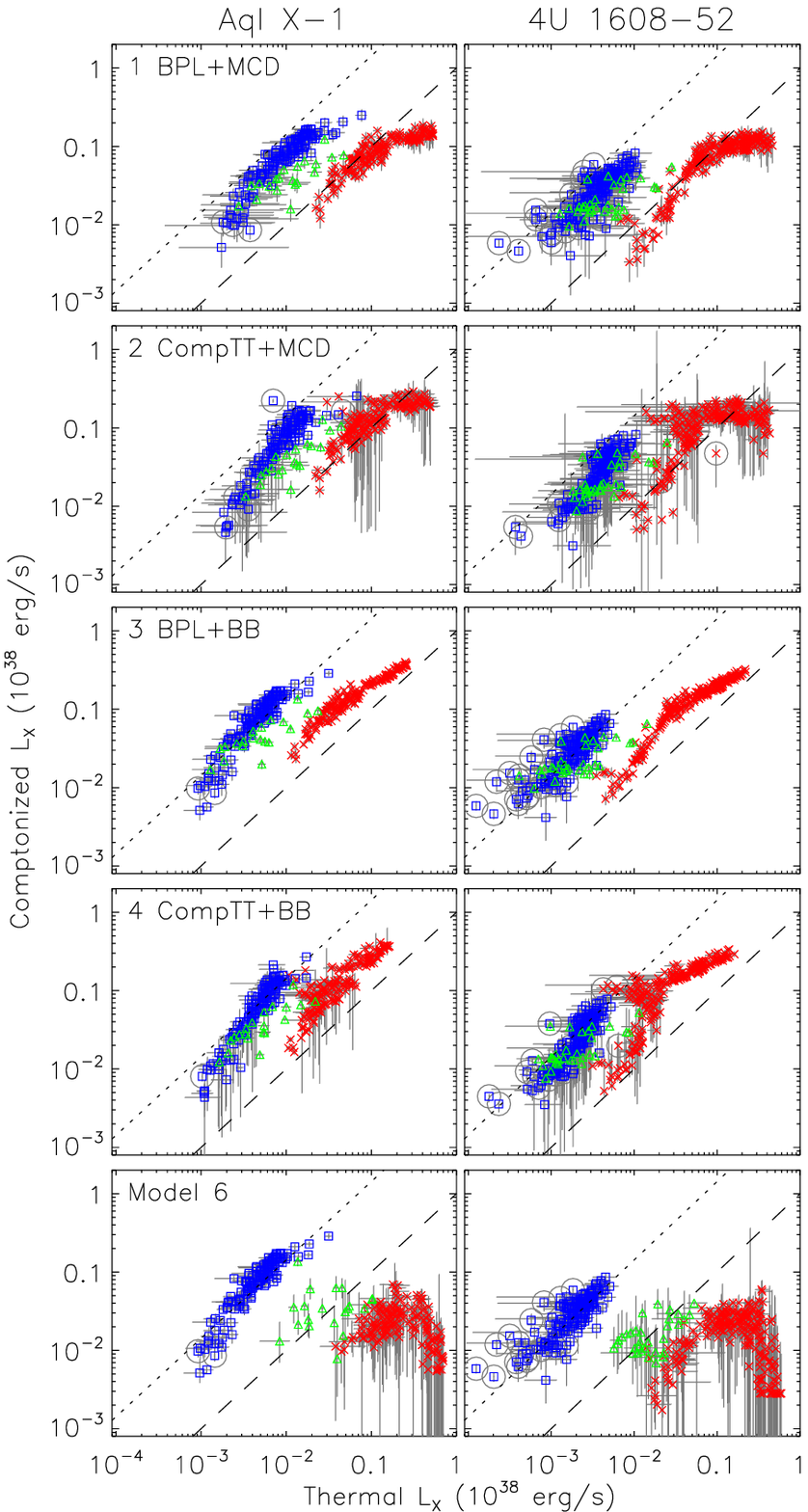}
\caption{The luminosity of the Comptonized component versus the
luminosity of the thermal component(s). The dashed and dotted lines
are for reference. They connect the points where the ratio of the
luminosities of the thermal component and the Comptonized component is
1.0 and 0.07, respectively. The latter is a typical value for the hard
state.
\label{fig:lumlum}}
\end{figure}

We fitted the observations of Aql X-1 and 4U 1608-52 using Models 1--4 in
Table~\ref{tbl-2}. The PCA does not allow us to gain meaningful
constraints on the properties of the (cool) thermal components using
the hot-seed-photon models, as explained above. Thus, for models with
CompTT, we just show results of cold-seed-photon solutions. In these
models, $kT_{\rm s} \ll kT_{\rm e}$ ($kT_{\rm e}$ is typically 2--3
keV in the soft state and several tens of keV in the hard state), and
$kT_{\rm s}$ is far below the PCA energy limit. Thus, the Wien approximation
of the input seed photon is valid.

The top four rows in Figures~\ref{fig:tcr}, \ref{fig:Tlum} and
\ref{fig:lumlum} show the fit results for the commonly used
Comptonized + thermal models. They correspond to Models 1, $2_{\rm
cold}$, 3, and $4_{\rm cold}$ in Table~\ref{tbl-2}, where we also list
the mean $\chi^2_\nu$ values. The errors represent 90$\%$ confidence
limits for a single parameter, and the circled points are those with
relatively large errors.  There is not only strong similarity between
these two sources, but also strong similarity between the results of
these four models. The extreme difference of the hard and soft states
can also be seen in these plots.

Figure~\ref{fig:tcr} shows the variation of the color temperature of
the thermal components ($kT_{\rm bb}$ for BB and $kT_{\rm mcd}$ for
MCD) with the PCA intensity. All four models show that the temperature
of the thermal components of both sources increases with intensity in
the hard state. In contrast, the temperature is almost constant (for
models with BPL) or decreases (for models with CompTT) with increasing
luminosity in the soft state.

Figure~\ref{fig:Tlum} shows the luminosity of the thermal component
versus its color temperature, $kT_{\rm bb}$ or $kT_{\rm mcd}$. For
reference, we also show the lines for constant radius, assuming
$L_{\rm X}=4\pi R^2\sigma T^4$. The NS radii (\S\ref{resmoddeg}) are shown
with dotted lines in this figure. The dashed lines correspond to $R =$
1.9 km and 1.3 km for \object{Aql X-1} and \object{4U 1608-52},
respectively; they are derived from the fit to the BB radius values
obtained from Model 6 (see below). With increasing luminosity in the
respective thermal component, $R_{\rm mcd}$ or $R_{\rm bb}$ increases
in the soft state.  For the case of MCD-related Models 1 and $2_{\rm
cold}$, this behavior in the soft state contradicts the basic
prediction of the accretion disk model and warrants the ``L'' problem
in Table~\ref{tbl-2}. Besides, from the comparison with the $R_{\rm
burst}$ lines, we find that the inner disk radius $R_{\rm mcd}$ in the
hard state and some part of the soft state is simply too small, even
after taking into account expected correction factors (see
\S\ref{sec:propertys}). This is consistent with the results of
\citet{gido2002b} and \citet{chba2001}. Thus, the ``R'' problem also
applies to these two models.
  
For BB-related Models 3 and $4_{\rm cold}$, the luminosity evolution
of the effective radius is tied to the evolution of the boundary
layer, which is much less certain. This is why the ``L'' problem
is not applied to the BB component in Table~\ref{tbl-2}. If we
attribute the thermal component BB to the boundary layer, it would
imply that the boundary layer is measured with almost constant surface
area in the hard state, but it spreads out with constant color
temperature in the soft state. One possible explanation for the constant color
temperature during the spreading of the boundary layer is that the
local flux reaches the Eddington limit.  The spreading layer model
\citep{supo2006} predicts that the color temperature of the boundary
layer is $\sim$ 2.4 keV for a NS with $R_{\rm ns} = 15$ km and $M =
1.4 M_\sun$ and solar composition of the accreting matter (smaller
$R_{\rm ns}$ would predict higher value). A problem with this
interpretation for our results of Model 3 or $4_{\rm cold}$ is that
the critical value of the temperature for area expansion is $\sim$ 1.5
keV. Higher temperatures are only seen during Type I X-ray bursts from
these two sources with maxima normally $> 2.5$ keV. Thus, the critical
value of the temperature $\sim$ 1.5 keV from Models 3 and $4_{\rm
cold}$ does not match expectations for boundary layer spreading. We will
show later that these two models are also disfavored from strong
similarities of the timing properties between BH and NS LMXBs.

In Figure~\ref{fig:lumlum} we investigate the luminosity evolution of
the Comptonized component (BPL or CompTT) versus the thermal
component(s) (BB or MCD or their sum if both are used). For
the thermal component the bolometric luminosity is used while for the
Comptonized component we integrated from 1 keV to 200 keV for CompTT
and from 1.5 keV to 200 keV for BPL. The 1.5 keV lower limit for the
BPL integration is chosen so that this component does not extend below
the temperature of the MCD in the soft state, i.e., when the BPL is
steep and the lower limit matters. The errors of the luminosity of
BPL, CompTT and MCD primarily depend on the uncertainties in their
respective normalizations. In Figure~\ref{fig:lumlum} we also show two
reference lines: dashed and dotted. They connect the points where the
ratio of the luminosities of the thermal component and the Comptonized
component is 1.0 and 0.07, respectively. The latter is a typical value
for the hard state. In both the hard and soft states,
the luminosity of the Comptonized component increases with the
luminosity of the thermal component, but the two states follow
different tracks.  In the soft state, the luminosities of these two
components are relatively close to each other, while in the
hard state, the luminosity of the thermal component is only $\sim
10\%$ of the Comptonized component.

The uncertainties for the thermal-component luminosity from the
hot-seed-photon models are quite large from PCA data in the soft
state. However, hot-seed-photon models with BB as the thermal
component have the ``T'' problem (\S\ref{moddeg}). Hot-seed-photon
models with MCD as the thermal component have the ``L'' problem in the
soft state \citep{gido2002b, dozysm2002} and the ``R'' problem in the
hard state from our investigation (not shown).

Since the luminosity evolution of the thermal component in the soft
state is either in violation of the basic model (MCD) or, at best,
suspicious (BB), we continued to investigate alternative models such
as those in the next section.

\subsection{Double Thermal Models}

\begin{figure} 
%\figurenum{9}  
\plotone{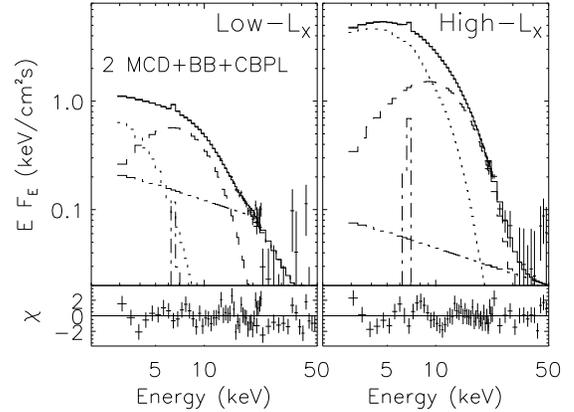} 
\caption{The unfolded spectra and residuals of two sample soft-state
observations of Aql~X--1 using Model 6. The total model fit is shown
as a solid line, the MCD component as a dotted line, the BB component
as a dashed line, the CBPL component as a dot-dot-dashed line, and the
Gaussian line as a dot-dashed line. The high-$L_{\rm X}$ observation
is the same as the sample soft-state observation seen in
Figures~\ref{fig:samspec} and \ref{fig:samspechot}.  The low-$L_{\rm
X}$ observations is on 1999 June 3 (hard color 0.32, intensity 243
counts/s/PCU).
\label{fig:samspecmod6}} 
\end{figure}

\begin{figure}
%\figurenum{10} 
\plotone{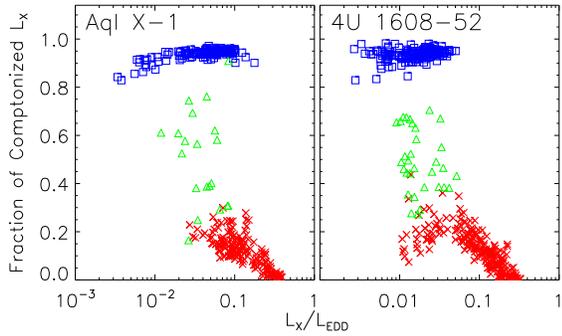}
\caption{The fraction of Comptonized luminosity versus the total
luminosity, using Model 6. $L_{\rm EDD}=1.8 \times 10^{38}$ erg/s, assuming $M_{\rm ns} =
1.4 M_\sun$ for both sources.
\label{fig:nonthermfraclum}}
\end{figure}

Early analyses of NS spectra in the soft state considered the
possibility that we might detect thermal components from both the disk
and boundary layer \citep{miinko1984}. This ``double thermal'' model
(i.e., MCD+BB or Model 5 in Table~\ref{tbl-2}) was applied to the
soft-state observations of \object{Aql X-1} and \object{4U
1608-52}, and the results are shown in the fifth row of panels in
Figures~\ref{fig:tcr} and \ref{fig:Tlum}. Data points for this model
are omitted when $\chi^2_\nu$ is $> 2$ (only for this model). It turns
out that this model works very well in the soft state when the
luminosity is high, but with the decreasing luminosity, $\chi^2_\nu$
becomes large. Either source gives a mean $\chi^2_{\nu}$ $\sim$ 1.2
and $\sim$ 3.5 for observations with source intensity $> 800$
counts/s/PCU and $< 500$ counts/s/PCU, respectively (also see
Table~\ref{tbl-2}). However, Model 5 is remarkably successful in its
implication that the disk and the boundary layer both remain at
constant sizes for a substantial range of luminosity
(Figure~\ref{fig:Tlum}). Moreover, the inferred size of the boundary
layer is close to that inferred by Models 3 and $4_{\rm cold}$ for the
hard state. The failure of Model 5 for soft-state observations at
lower luminosity is apparently due to small levels of Comptonization,
since the fit residuals are pointing chronically positive at photon
energies above 15 keV (PCA data).

Given the interesting luminosity evolution for the soft state with
Model 5 (no Comptonization), we added a {\it weakly Comptonized} component, while
continuing to assume that both the MCD and BB are visible in the soft
state. The difficulty here is how to model weak Comptonization by
adding a third component. Freely adding a BPL or
CompTT component is obviously not feasible, because BPL or CompTT plus
one thermal component is sufficient to model the entire spectrum as
shown by Models 1--4. As a first attempt, we tried several forms
of constrained CompTT (e.g., couple $kT_{\rm s}$ to $kT_{\rm bb}$ or
$kT_{\rm mcd}$; force $kT_{\rm e}> 10$ keV; or do both), but the fits
always yielded large fractions of Comptonization for low-luminosity
observations in the soft state. This problem could be due to RXTE's
lack of low-energy coverage.

On the other hand, we found that the entire soft state remains
weakly Comptonized if we adopt the following constrained BPL (CBPL)
model: the break energy is fixed at 20 keV (the best-fitting break
energy is typically $> 20$ keV in the hard state and is $\sim 15$ keV
in the soft state from Models 1 and 3), and the initial photon index
is required to be $\le 2.5$ \citep[a typical initial photon index in
the soft state of BHs is also about 2.5;][]{remc2006}. Therefore, we
define Model 6 as follows: the hard state is still modeled by BPL+BB
(no BPL constraints; same as Model 3 for the hard state) and the
soft/transitional states are modeled by MCD+BB+CBPL.
Figure~\ref{fig:samspecmod6} shows the unfolded spectra of two
soft-state observations using this model, one at low luminosity and
the other at high luminosity.

The bottom row of panels of Figures~\ref{fig:tcr}, \ref{fig:Tlum} and
\ref{fig:lumlum} show the results obtained for Model 6.
Figure~\ref{fig:tcr} shows that with the increase in the intensity,
$kT_{\rm bb}$ increases both in the hard and soft states, with the
tracks that are clearly separated. The BB temperature reaches a
maximum $\sim$ 2.5 keV for \object{Aql X-1} and $\sim$ 3.0 keV for
\object{4U 1608-52}, similar to the peak temperatures of Type
I X-ray bursts from these two sources \citep{koinma1981, nadoin1989,
gamuha2006}. The temperature at the inner disk radius $kT_{\rm mcd}$
increases from $\sim$ 0.5 to 2.0 keV, also correlated with the
intensity.

In Figure~\ref{fig:Tlum}, we see that Model 6 remarkably produces
results where $L_{\rm X} \propto T^4$ for both the MCD and BB components in
the soft state.  Furthermore, the inferred emission areas of the
boundary layer from the hard and soft states have essentially the same
value. The emission area of the boundary layer is small compared with
the size of the NS. We realize that the size of the BB emission area
slightly increases with decreasing luminosity in the hard state,
especially for \object{4U 1608-52}. However, we note that the BB curve
lies well below the line of $R_{\rm burst}$ at all luminosities.

In Figure~\ref{fig:lumlum}, we show the luminosity of the Comptonized
component versus the total luminosity of the double thermal
components. The behavior in the soft state for Model 6 is quite
different from that in the first four models. Here, with increasing
luminosity of the thermal components (i.e. $L_{\rm bb}+L_{\rm mcd}$),
the luminosity of the Comptonized component first increases and then
decreases. At the highest luminosities, the Comptonized component is
negligible, as implied by the success of Model 5 in the same region.
Figure~\ref{fig:nonthermfraclum} shows the fraction of the Comptonized
luminosity versus the total luminosity. It looks similar to the
color-intensity diagram in Figure~\ref{fig:ccdiag}, suggesting that
the hard color tracks the degree of Comptonization fairly well.

It should be noted that there are other kinds of weak-Comptonization
approximations like constrained power-law (photon index $< 2.5$) or
constrained cutoff power-law (photon index $< 2.5$ and cutoff energy
$> 10$ keV) that gave results for the soft state that are similar to
those obtained using CBPL. For the hard state, Figures 6--8 show that
the Model $4_{cold}$ (CompTT+BB) gives similar results to Model 6 in
terms of the properties of the thermal component and the fraction of
Comptonization. Model 6 is successful, but there are no claims that it
is either a unique solution to the problem, nor an adequate depiction
of Comptonization other than the estimate for the fraction of energy
related to Comptonization.

\section{TIMING PROPERTIES AND COMPARISON WITH BLACK HOLES}
\label{sec:combh}

\begin{figure}
%\figurenum{11}
\plotone{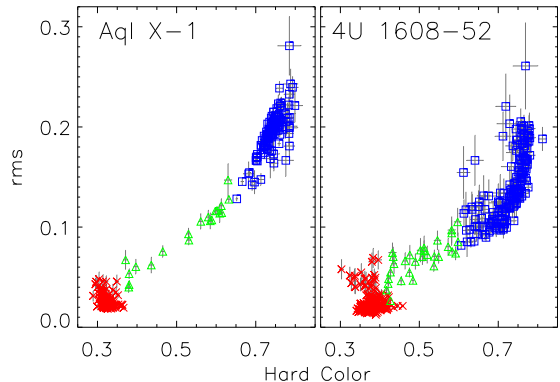}
\caption{The integrated rms power in the power density spectrum
(0.1--10 Hz and energy band 2--40 keV) versus the hard color.
\label{fig:rmshc}}
\end{figure}

\begin{figure}
%\figurenum{12}
\plotone{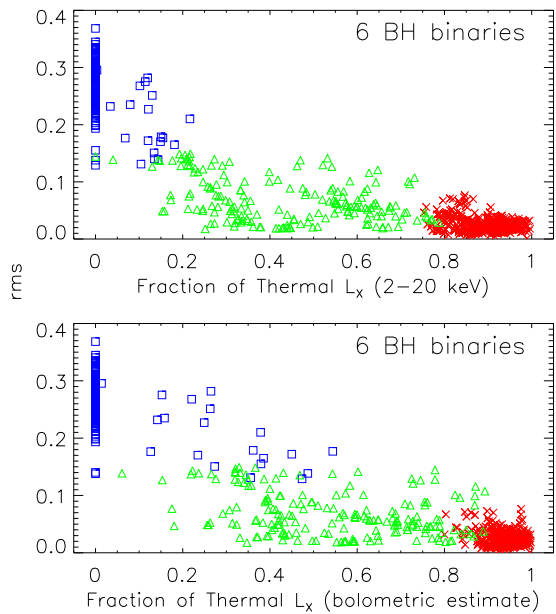}
\caption{The integrated rms power versus the fraction of luminosity 
contained in the thermal (MCD) component for 6 black hole systems.
\label{fig:rmslumfracbh}}
\end{figure}

\begin{figure}
%\figurenum{13}
\plotone{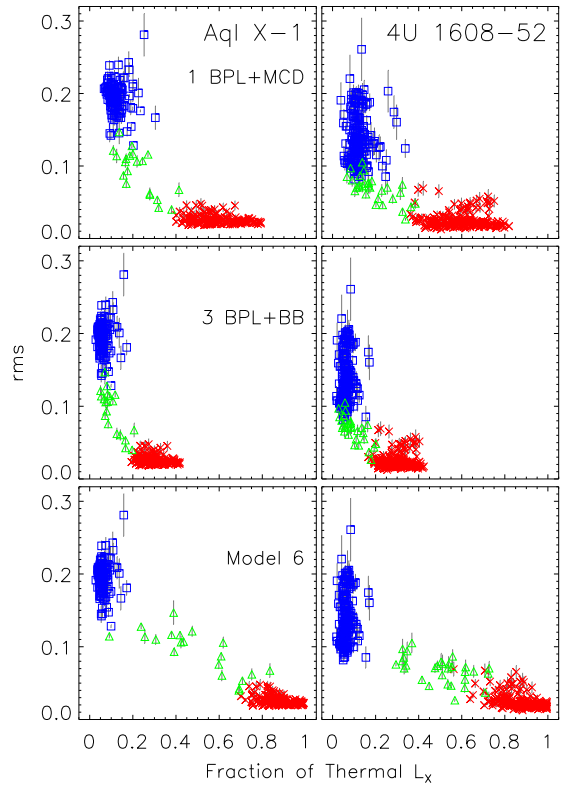}
\caption{the rms versus the fraction of luminosity contained in BB
and/or MCD components as evaluated for different spectral models.
\label{fig:rmslumfrac}}
\end{figure}

Compared to Models 1--4, the use of a double thermal model + CBPL
(i.e., Model 6) for \object{Aql X-1} and \object{4U 1608-52} produces
dramatically different results for the luminosity evolution of the
thermal components in the soft state and for the implied significance
of Comptonization in the soft state.

There are strong similarities in timing properties between the BH and
NS LMXBs (see, e.g., \citet{wi2001}), which can be used to further
assess these differences. In Figure~\ref{fig:rmshc}, we show the
integrated root-mean-square (rms) power in the power density spectrum
(0.1--10 Hz and energy band 2--40 keV) versus the hard color for
\object{Aql X-1} and \object{4U 1608-52}. The two sources show very
similar timing properties.  The rms, normalized to a fraction of the
source's mean count rate, is very small ($\lesssim0.05$) in the soft
state, and the values increase with the hard color progressing through
the transitional and hard states. The rms versus hard-color relations
from Figure~\ref{fig:rmshc} can be compared with similar plots for BHs
in \citet{remc2006} (panels g in their Figures 4--9). Based on such a
comparison we conclude that, when considered in a model-independent
manner, the way in which the strength of the X-ray variability changes
as a function of spectral hardness is very similar for BHs and our two
NS transients. Since the hard color effectively traces the fractional
contribution of the thermal and Comptonized components to the X-ray
spectrum in BH systems, it is interesting to see which of our spectral
models conserves this similarity when rms is plotted versus the
fractional contribution of the thermal component(s).

In Figure~\ref{fig:rmslumfracbh} we present such a plot for the same 6
BH binaries analyzed by \citet{remc2006}. Two different integration
limits were investigated for the thermal and Comptonized luminosities,
but the overall behavior was insensitive to our exact choice. It can
be seen that for BHs in the soft state, the luminosity fraction of the
thermal component is quite high ($\gtrsim75\%$) and the rms is very
small $\lesssim0.05$. For \object{Aql X-1} and \object{4U 1608-52} we
plot the rms versus the fraction of thermal luminosity for Models 1,
3, and 6 in Figure~\ref{fig:rmslumfrac}.  Note that in case of Model 6
the sum of MCD and BB was used to determine the fractional
contribution of the thermal components. Comparing Figures
\ref{fig:rmslumfracbh} and \ref{fig:rmslumfrac} one can see that only
Model 6 reproduces the same dependence of rms on fractional thermal
luminosity for \object{Aql X-1} and \object{4U 1608-52}, as for the
six BHs. In Models 1 and 3, the luminosity fraction of the thermal
component can be as low as 40$\%$ and $20\%$, respectively, while
continuing to show very low values of rms power. Similar results are
obtained for Models 2 and 4 (cold and hot included for both
cases). Hence, for the models outlined in Table~\ref{tbl-2} only Model
6 links weak Comptonization with low rms power, as is clearly evident
in the properties of BHs. Models 1--4 for the soft state are therefore
described as having the ``P'' problem (see \S\ref{resmoddeg} and
Table~\ref{tbl-2}).

\section{PHYSICAL INTERPRETATIONS OF MODEL 6 FOR THERMAL COMPONENTS}
\label{sec:propertys}

Since Model 6 has many attractive advantages over the other models, we
further explore its implications in terms of the physical properties
of the accretion flow in both the hard and soft states.

There is a well known difficulty in deriving true radii from the
apparent dimensions (i.e., $R_{\rm bb}$, $R_{\rm burst}$, and $R_{\rm
mcd}$ extracted from model components of the X-ray spectrum BB,
BB$_{\rm burst}$, and MCD, respectively). Nevertheless, the small
value of $R_{\rm bb}$ and the nearly constant values of $R_{\rm bb}$
and $R_{\rm mcd}$ across a large range in luminosity motivate cautious
efforts to discuss physical interpretations.

If we assume the BB emission area to be a latitudinally symmetric
equatorial belt, then the BB emission area of the belt should be
\begin{equation}
A_{\rm belt} = 4\pi N_{\rm bb} D^2_{10 \rm kpc} f^4 k(i,\delta)  \ {\rm km}^2\label{eq:1},
\end{equation}
where $N_{\rm bb}=R^2_{\rm bb}/D^2_{10 \rm kpc}$ is the normalization
of the BB component (isotropic assumption), $f$ represents a spectral
hardening factor (sometimes expressed as the ratio of the color to
effective temperature), and $k$ is a geometrical correction factor
taking into account the emitting geometry and any occultations by the
accretion stream. The $k$ factor depends on the disk inclination
($i$), the latitude range (from the NS equator) of surface emission
($\delta$). We note that this factor is also strongly dependent on the
properties of the occulting accretion stream. Unless otherwise
indicated, we assume a geometrically thin but optically thick
accretion stream so that the half belt on the other side of the
accretion stream is invisible to the observer. To have some sense of
this $k$ factor, we give two examples: $k(i=60\degr,\delta=10\degr) =
1.69$, and $k(i=60\degr,\delta=90\degr) = 1.33$.  From this
perspective, the area of the NS, e.g., for \object{Aql X-1} in this
study is:
\begin{equation}
A_{\rm ns} = 4\pi N_{\rm burst}  D^2_{10 \rm kpc} f^4 k(i,\delta=90\degr) \ {\rm km}^2,
\end{equation}
assuming that the entire NS surface is radiating (using the asymptotic
value of $R_{\rm burst}$ late in the burst). This expression conveys
the difficulty in gaining accurate inferences of NS sizes from X-ray
burst measurements. The raw values for $R_{\rm burst}$ given in
\S\ref{resmoddeg} ignore $f$ and $k$.

Perhaps of greater interest is the effort to understand a key
measurement result of this paper: $R_{\rm bb} / R_{\rm burst} \simeq
0.25$ ($N_{\rm bb} / N_{\rm burst} \simeq 1/16$) for \object{Aql
  X-1}. If we assume that the $f$ values cancel in these different
applications of the BB model and use the fact $A_{\rm belt} / A_{\rm
  ns}=\sin\delta$, we obtain
\begin{eqnarray}
\sin\delta &=& \frac{N_{\rm bb} k(i,\delta)}{N_{\rm burst} k(i,\delta=90\degr)}=\frac{k(i,\delta)}{16k(i,\delta=90\degr)}.
\end{eqnarray}
In this case, the $R_{\rm bb} / R_{burst}$ value is equivalent to an
equatorial belt with a half-angle of $\delta_{\rm belt} \simeq 9\degr$
for $i = 30^\circ$, or $\delta_{\rm belt} \simeq 6\degr$ for $i =
60^\circ$.  We stress that this scaling estimate ignores the annular
half width of the occulting stream ($\delta_{\rm stream}$) itself. A more
realistic estimate is $\delta_{\rm belt}\gtrsim\delta_{\rm
stream}+6\degr$ for $i=60\degr$, implying $k \gg 1$.

Regardless of the details of $\delta$, we must confront the
implication of measuring $N_{\rm bb}/N_{\rm burst}\sim 1/16$, noting
that $N_{\rm burst}$ is the asymptotic value that should screen out
effect of the radius expansion and momentary disruption in the
occulting stream (\S\ref{resmoddeg}). How is it possible that
accreting NSs could maintain an almost uniformly small area of BB
emission through hard and soft states that span such a wide range in
luminosity (i.e., 0.005 to 0.5 $L_{\rm EDD}$)?  Such simple results
suggest a picture in which a geometrically thin accretion stream feeds
a rather well-defined impact zone, where the accreting gas may
efficiently release energy before spreading over the remainder of the
NS surface. Since the scale height of the inner accretion disk is
expected to increase with the accretion rate, it is difficult to
understand our results without the help of an inner-most stable
circular orbit (ISCO). As illustrated for BHs, the effective potential
of an ISCO creates a pinch on the vertical structure of the accretion
stream that can suppress variations in the scale height of the inner
disk \citep{abjasi1978}. To the extent that Model 6 remains viable
with further scrutiny, the small size of $R_{\rm bb}$ over a large
range in luminosity should be examined as a means to infer that these
NSs lie within their ISCOs. This may provide another observational
link to general relativity, while providing potential constraints on
the NS equation of state.

The ability of Model 6 to restore the expected luminosity evolution of
the MCD component motivates efforts to compare the MCD radius and
relative luminosity to the results determined for the BB component.
In Figure~\ref{fig:Tlum}, it is evident that the raw values of $R_{\rm
mcd}$ are only slightly less than the values of $R_{\rm burst}$ (dotted
line) for \object{Aql X-1} and \object{4U 1608-52}. Given the
uncertainty in the different correction factors that must be applied,
respectively, to the disk and the burst radii in order to derive
physical sizes, our results may still be consistent with expectations
framed by the preceding discussion, i.e., that $R_{\rm ns} < R_{\rm
isco} \lesssim R_{\rm disk}$.

The comparison of BB and MCD luminosities is a trickier topic.  The
apparent luminosity of the boundary layer ($L_{\rm bb}$) is about one
third of that of the disk ($L_{\rm mcd}$) in the soft state,
neglecting the weak Comptonization. In contrast, we might expect
$L_{\rm bb} \gtrsim L_{\rm mcd}$ if the boundary layer is inside the
NS ISCO, as sketched above, where the accreting material may impact
the surface with a large relative velocity that includes a component
along the radial path.  We note, however, that while $L_{\rm mcd}$ is
estimated in a true bolometric sense, this is not the case for $L_{\rm
bb}$. We need to correct $L_{\rm bb}$ by $k(i,\delta)$
(Equation~\ref{eq:1}) due to the geometry of the boundary layer and
occultation by the accretion stream. For \object{Aql X-1}, it is
$\simeq 1.7$ ($i=60\degr$) or $\simeq 2.6$ ($i=30\degr$), and these
factions can become much larger if $\delta_{\rm stream}\sim
\delta_{\rm belt}$, as noted above.  We conclude that there is
considerable uncertainty in our final results as to whether the total
energy losses at the boundary layer are less than the total bolometric
luminosity of the other spectral components.

\section{PHYSICAL CONSEQUENCES OF MODEL 6 FOR THE HARD STATE}
\label{sec:propertyh}

\begin{figure}
%\figurenum{14} 
\plotone{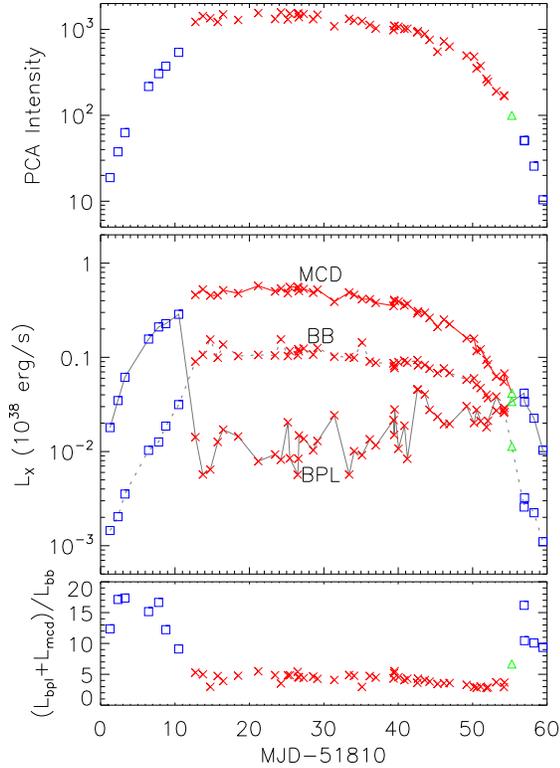}
\caption{The luminosity evolution of different spectral components
(middle panel) during the 2000 outburst of Aql~X--1, as viewed with
Model 6.  Blue, green, and red symbols denote hard, transitional, and
soft states, as done previously.  Note that the MCD component (points
connected by a solid red line) is not directly visible with the PCA in
the hard state.  The top panel shows the PCA count rate (2-30 keV).
The bottom panel shows the ratio: ($L_{\rm bpl} + L_{\rm mcd}$) /
$L_{\rm bb}$, and the value is clearly highest in the hard state, when
Comptonization appears to dominate the X-ray spectrum.
\label{fig:aqlx12000}}
\end{figure}

\begin{figure}
%\figurenum{15}
\plotone{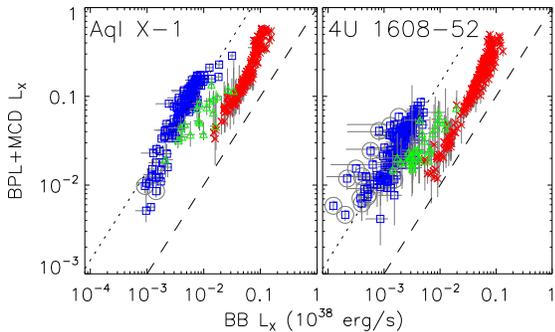}
\caption{The luminosity of non-BB components (i.e., BPL+MCD) versus
the luminosity of the BB component. At the transitional BB luminosity
that separates hard and soft states, the hard-state BPL appears to be
very luminous compared to the soft-state MCD. The two reference lines,
dashed and dotted lines, connect the points where the ratio of the
luminosities of the BB component and the non-BB components is
1.0 and 0.07, respectively.
\label{fig:pldiskbb}}
\end{figure}

Figure~\ref{fig:aqlx12000} shows the luminosity evolution with time
during a well-covered outburst of \object{Aql X-1} in 2000. The light
curve of this outburst is shown on the top panel. The middle panel
shows the luminosity evolution of each spectral component, using our
hybrid Model 6, where the BPL and MCD dominate the apparent luminosity
of the hard and soft states, respectively.  The bolometric BB
luminosity ($L_{\rm bb}$) curve resembles the total luminosity curve,
and it provides a reference measurement that allows us to further
compare the hard and soft states.  The ratio of BPL (isotropic) and
MCD (bolometric) luminosities ($L_{\rm bpl} + L_{\rm mcd}$) to $L_{\rm
bb}$ is shown in the bottom panel. We note that the values of this
ratio are elevated, in part, by the $k$ factor described for
Eq. 1. Nevertheless the high luminosity of the hard state, relative to
$L_{\rm bb}$, is apparent in the bottom panel of
Figure~\ref{fig:aqlx12000}.

To compare the hard and soft state luminosities in a more global
manner, Figure~\ref{fig:pldiskbb} shows ($L_{\rm bpl} + L_{\rm mcd}$)
versus $L_{\rm bb}$ for all of the \xte observations of \object{Aql
X-1} and \object{4U 1608-52}. The behavior of these two sources is
quite similar.  The states are largely separated by vertical lines
near (uncorrected) $L_{\rm bb} \sim 10^{36}$ erg/s.  Here, the
hard-state values for ($L_{\rm bpl} + L_{\rm mcd}$) are clearly higher
than those for the soft state, by a factor of $\sim 6$.  For the hard
state, the vertical axis represents Comptonization, since $90\%$ of
the total luminosity in the hard state is due to $L_{\rm bpl}$.
In the soft state, the vertical axis is effectively the disk
luminosity. On the horizontal axis, we assume that $L_{\rm bb}$
represents emission from the boundary layer, and we can then use the
nearly constant value of $R_{\rm bb}$ across the hard and soft states
to motivate the presumption that $L_{\rm bb}$ quantifies the mass
accretion rate onto the NS surface.  Using these assumptions, there
are still different ways to interpret Figure~\ref{fig:pldiskbb}.

First, if we further assume that the total accretion rate (\mdot)
flows into a boundary layer with constant $\delta_{\rm belt}$, then
$L_{\rm bb}$ effectively tracks (\mdot), and we can compare hard and
soft states in terms of their radiation efficiency. Unlike $L_{\rm
bb}$, $L_{\rm mcd}$ and $L_{\rm bpl}$ are not expected to require
large correction factors, although the origin of the BPL is very
uncertain.  We would then conclude from Figure~\ref{fig:pldiskbb} that
Comptonization in the hard state has higher radiative efficiency than
thermal disk emission in the soft state, and this occurs across a wide
range in the accretion rate. This idea is a radical departure from
expectations.

Alternatively, we can refrain from making assumptions about \mdot, and
choose a line of constant ($L_{\rm bpl} + L_{\rm mcd}$) in
Figure~\ref{fig:pldiskbb}.  Quite generally, then, one would conclude
that less mass reaches the NS surface in the hard state, compared to
the soft state, for a given level of luminosity on the vertical axis.
This might suggest that the majority of the mass flow is disrupted
from the normal path to the NS in the hard state.  Given the
expectation for radio jets in the hard state of accreting NS LMXBs
\citep[e.g.,][]{fe2006,mife2006}, this could further imply that a
significant fraction of mass accretion in the hard state is redirected
to outflow in the jet. The energy source for this scenario could be
problematic.

Finally, it is also possible that some of our preliminary assumptions
are incorrect.  Changes in accretion geometry across the state
boundary, e.g., from correlated jumps in $\delta_{\rm belt}$ and
$\delta_{stream}$, would be one way to invalidate the assumption that
$L_{\rm bb}$ uniformly tracks the NS surface accretion rate across both
states. All of these ideas need to be investigated for other sources
and in different ways.

\section{SUMMARY AND DISCUSSION}

We have analyzed ten years of \xte observations of two NS
X-ray transients, \object{Aql X-1} and \object{4U 1608-52}, through
more than twenty individual outbursts.  Although there is much variety
in these outbursts, the spectral and timing properties are well
organized, as has been demonstrated in previous investigations
\citep{murech2002, gido2002a}. It is also well known that several
different models may adequately fit the X-ray spectra of NS LMXBs,
while each one may convey very different physical implications, e.g.,
the temperatures and sizes of the emitting regions. We therefore
attempted to design performance-based criteria for evaluating these
models, hoping to gain insights on a model's self consistency, while
taking advantage of the large data archive for these two
atoll-transient prototypes.

Among these criteria, there is an important quantity, which is the
energy fraction for Comptonization, i.e., how much energy is directly
visible as pure thermal radiation and what remaining fraction is
diverted and expressed as inverse Compton radiation.  This quantity is
interesting from an accretion-energetics point of view. It can be
calculated from any of these models, and furthermore, it is the
gateway for comparing NSs and BHs in terms of the relationship between
energetics and the measured timing properties.

First we examined the classical two-component models with a
Comptonized component (BPL or CompTT) and a thermal component (MCD or
BB). We confirmed that many combinations can describe the X-ray
spectra successfully. Moreover there are two possible seed-photon
solutions for models that utilize CompTT, a problem that is also
present for instruments with extended low-energy coverage like {\it
BeppoSAX}. In the soft state, the two-component models behaved poorly
for a variety of reasons shown in Table~\ref{tbl-2},
Figure~\ref{fig:Tlum}, and
Figures~\ref{fig:rmslumfracbh}--\ref{fig:rmslumfrac}. We then
progressed to consider a double-thermal model (MCD+BB) for the soft
state, and we found a need to account for weak Comptonization at the
low-$L_{\rm X}$ end of the soft-state track. We then devised a hybrid
``Model 6'' in which the hard state is modeled by BB+BPL while the
soft state is modeled by MCD+BB+CBPL, where the third component is a
constrained BPL ($E_{\rm b} = 20$ keV and $\Gamma_1 \leq 2.5$) that
can assume the role of weak Comptonization.

It turns out that Model 6 offers three great advantages over the
classical Comptonized + thermal Models. (1) It produces an $L_{\rm X}
\propto T^4$ relation for both the MCD and BB in the soft state with a
sufficiently large value for the inner disk radius (compared to the NS
radius inferred from Type I bursts). (2) The emission area of the BB
extends from the hard to the soft state with essentially the same
value. (3) The fraction of Comptonization inferred by this model is
also consistent with that for BHs for given values of rms power in the
power density spectra.  When BH binaries are in the thermal state,
the Comptonization fractions in the energy spectra are very low and
the rms power is also very low, rms $< 0.05$. Atoll sources in the
soft state show similar rms values, but only Model 6 implies low
Comptonization for the soft state.

With regard to previous investigations of NS X-ray spectra (\S1), our
Model 6 splices together the Western model for the hard state (BB + a
heavily Comptonized disk) and a three-component model for the
transitional and soft states (BB + MCD + constrained BPL) that
represents a BB plus a very weakly Comptonized disk.  The soft-state
portion of this model was heavily influenced by the original version
of the Eastern model (Mitsuda et al. 1984). We further note that low
Comptonization solutions ($\tau \sim 1$; $kT_{\rm e} > 20$ keV) for
the soft state were derived with a two-component Eastern model for
\object{4U 1608-52} with {\it Tenma} data (Mitsuda et al. 1989), and these
are the closest results that we can find to those (soft state) of
Model 6.

The ad hoc scheme in Model 6 to handle weak Comptonization with the
CBPL component for the soft state is obviously a topic that needs
further work. Model 6 for the soft state is a three-component
continuum model. We did not try to test three-component models in an
open-handed manner. Instead, we focused on low-Comptonization
solutions, progressing from the success of Model 5 (no
Comptonization). The CBPL used in this paper is just one possible
solution; our investigation showed that constrained power-law with
photon index $< 2.5$ or constrained cutoff power-law with photon index
$< 2.5$ and cutoff energy $> 10$ keV would also produce similar
results to those obtained with the CBPL. As other alternatives, we
tried to add a constrained version of CompTT to the two thermal
components, by coupling $kT_{\rm s}$ to $kT_{\rm bb}$ or $kT_{\rm
mcd}$, or requiring $kT_{\rm e} > 10$ keV, or doing both. However, the
fits always yielded very strong Comptonization in the low-luminosity
soft state, but the fit residuals for Model 5 suggested that only weak
Comptonization is present in the low-luminosity soft state. This could
be due to the limited low-energy coverage of RXTE data. We
investigated {\it BeppoSAX}, which has extended low-energy coverage,
for possible new insight. Unfortunately, each X-ray source has only
one soft-state observation from {\it BeppoSAX}, and both of them are
at high luminosity. Future observations with broad-band instruments
like {\it Suzaku} could be important to this issue. As for the hard
state, Model 6 uses BB+BPL. Replacement of BPL with CompTT does not
change our conclusions for the properties of the BB component and the
fraction of Comptonization as long as the cold-seed-photon solution is
used.

There is a question of whether we could also see a thermal disk in the
hard state, where the trend from Figure~\ref{fig:lumlum} would imply
$kT_{\rm mcd} \lesssim$ 0.5 keV. Such disks have in fact been observed
in the hard state of a number of BH X-ray transients
\citep{mihomi2006,mihost2006,rymist2007}, with instruments that have
better sensitivity at low energies than RXTE. However, being able to
detect these components in NSs is not only a matter of low-energy
sensitivity. One also needs a relatively low interstellar
absorption. Moreover, in NSs a disk would not be the only thermal
component contributing around 0.5 keV, since there is also emission
from the boundary layer, and it is difficult to disentangle the
contribution from the two components at low luminosities when the
spectrum is not dominated by these thermal components. 

If the boundary layer is an equatorial belt of BB emission, then Model
6 implies that the visible surface area of the belt remains nearly
constant, with $N_{\rm bb}/N_{\rm burst} \sim 1/16$, over a wide range
in $L_{\rm X}$ (0.005--0.5 $L_{\rm Edd}$). Current theories predict
that the geometry and physical processes in the boundary layer should
vary significantly with $L_{\rm X}$ \citep[e.g.,][]{insu1999,posu2001,
klwa1985}, and some observations have been interpreted in such a
manner \citep{chinba2002}. Our results are much simpler than expected,
suggesting that a geometrically thin accretion stream impacts a rather
well-defined surface area, where the gas radiates efficiently before
spreading over the NS. This scenario would seem to require that the NS
lies within its ISCO, which pinches the vertical structure of the
accretion stream \citep{abjasi1978} as it flows toward the NS.  This
topic must be investigated in further detail.

\acknowledgements We thank Marek Abramowicz, Wlodek Klu{\'z}niak,
Ramesh Narayan, Jeff McClintock, Lev Titarchuk, and Miriam Krauss for
very helpful discussions. Primary support for this research was
provided by the NASA contract to MIT for RXTE instruments.

\bibliographystyle{aa}
\bibliography{all-bib}

\clearpage

\begin{deluxetable}{ccccc}
\tabletypesize{\scriptsize}
\tablecaption{X-ray Sources and Observations Prior to 2006 January 1 \label{tbl-1}}
\tablewidth{0pt}
\tablehead{
 &  \colhead{number of total/used} & \colhead{Time of total/used} &  \colhead{{\it N}$_{\rm H}$} &\colhead{Distance}\\
\colhead{Source name} & \colhead{observations} & \colhead{observations (ks)} & \colhead{($10^{22} {\rm cm}^{-2}$)} &\colhead{(kpc)}
}
\startdata
\object{Aql X-1} & 393/333 &  1402/1252 &  0.5 & 5\\  
\object{4U 1608-52} & 523/459 & 1310/1204 & 1.0 & 3.6\\
\enddata
\end{deluxetable}

\begin{deluxetable}{lccccccc}
\tabletypesize{\scriptsize}
\tablecaption{The spectral models\label{tbl-2}}
\tablewidth{0pt}
\tablehead{
& \multicolumn{3}{c}{Hard State} & &  \multicolumn{3}{c}{Soft/Transitional State}\\
\cline{2-4} \cline{6-8} \\
 \colhead{Model \#\tablenotemark{a}} & \colhead{Description} & \colhead{$\chi^2_{\nu}$($\sigma$)\tablenotemark{b}} & \colhead{Problems\tablenotemark{c}} & & \colhead{Description}& \colhead{$\chi^2_{\nu}$($\sigma$)\tablenotemark{b}} &  \colhead{Problems\tablenotemark{c}}
}
\startdata
1  & BPL+MCD & 0.93(0.21),0.88(0.16) & R & & BPL+MCD &0.99(0.38),1.07(0.36) &  R L P\\  
$2_{\rm cold}$  & CompTT+MCD &1.00(0.28),0.90(0.17) & R & & CompTT+MCD &1.04(0.39),0.93(0.25)& R L P\\  
$2_{\rm hot}$  &  CompTT+MCD & ... & R & & CompTT+MCD & ... &  L P \\
3  & BPL+BB &0.94(0.21),0.90(0.17) & --- & & BPL+BB &1.06(0.41),1.24(0.49) &  P\\  
$4_{\rm cold}$  & CompTT+BB &0.99(0.24),0.93(0.19) & --- & & CompTT+BB &0.95(0.33),0.90(0.25) &  P \\  
$4_{\rm hot}$  &  CompTT+BB & ... & T  & & CompTT+BB & ... & T P \\
5\tablenotemark{d} & ... & ... &... & & MCD+BB & 1.23(0.36),1.17(0.38) &  ...\\  
6 & BPL+BB &0.94(0.21),0.90(0.17) & --- & & MCD+BB+CBPL &1.15(0.30),1.02(0.27) & ---\\  
\enddata

\tablecomments{All models also include an interstellar absorption
component and a Gaussian line.  The notations BPL, MCD, BB and CompTT
refer the bknpower, diskbb, bbodyrad, and comptt models in XSPEC, respectively.
CBPL is a constrained BPL with the break energy $E_{\rm b}$ fixed at $20$ keV
and the initial photon index $\Gamma_1$ forced to be smaller than $2.5$. ``...'' means that
the information is not available or meaningless because the
corresponding model does not work. ``---'' means that none of the
four problems listed in \S\ref{resmoddeg} is found to apply to the
corresponding model/spectral state.}

\tablenotetext{a}{The subscripts cold and hot denote the cold-seed-photon models and hot-seed-photon models, respectively.}
\tablenotetext{b}{The mean $\chi^2_{\nu}$ and standard deviation. The two columns are for Aql~X--1 and 4U~1608--52, respectively.}
\tablenotetext{c}{See \S\ref{resmoddeg} for the meanings of these characters.}
\tablenotetext{d}{Only  observations in the soft state are fitted with this model. The mean $\chi^2_{\nu}$ for this model is from observations with source intensity $> 800$ counts/s/PCU.}
\end{deluxetable}

\begin{deluxetable}{llcccccc}
\tabletypesize{\scriptsize}
\tablecaption{BEST-FITTING PARAMETERS OF TWO SAMPLE SPECTRA \label{tbl-3}}
\tablewidth{0pt}
\tablehead{
\colhead{Data} & \colhead{Model} &\colhead{$kT_{\rm s}$ keV} & \colhead{$kT_{\rm e}$ keV} & \colhead{$\tau$} & \colhead{$kT_{\rm mcd},kT_{\rm bb}$ keV} &  \colhead{$N_{\rm mcd},N_{\rm bb}$} &\colhead{$\chi^2_\nu$(d.o.f)}
}
\startdata
Hard-state sample & CompTT+MCD(cold) & $\lesssim0.35$           & $14.3_{-0.9}^{+1.0}$ & $6.07_{-0.36}^{+0.36}$ & $1.79_{-0.08}^{+0.07}$ & $ 1.7_{-0.3}^{+0.3} $ & $1.49 (78)$\\
Hard-state sample & CompTT+MCD(hot) & $1.16^{+0.08}_{-0.12}$ & $15.5^{+1.2}_{-1.0}$ & $5.54^{+0.32}_{-0.32}$ & $1.00^{+0.05}_{-0.09}$ & $ 44.8^{+17.2}_{-10.7}$ & 1.32(78)\\  
Hard-state sample & CompTT+BB(cold) & $\lesssim0.4$             & $15.0_{-0.9}^{+1.0}$ & $2.53_{-0.14}^{+0.14}$ & $1.21_{-0.04}^{+0.04}$ & $ 8.9_{-1.5}^{+1.7} $ & $1.25 (78)$\\
Hard-state sample & CompTT+BB(hot) & $1.03_{-0.18}^{+0.12}$  & $16.6_{-1.1}^{+1.3}$ & $2.24_{-0.14}^{+0.14}$ & $0.67_{-0.05}^{+0.05}$ & $ 227.9_{-44.0}^{+63.1} $ & $ 1.59 (78 )$\\
Soft-state sample & CompTT+MCD(cold) & $\lesssim0.5$            & $2.5_{-0.1}^{+0.1} $ & $16.6_{-2.5}^{+8.5} $  & $1.70_{-0.06}^{+0.08}$ & $ 68.7_{-15.7}^{+22.6} $ & $ 1.27 (53)$ \\
Soft-state sample & CompTT+MCD(hot) & $1.02_{-0.03}^{+0.02}$ & $2.6_{-0.1}^{+0.1}$  & $10.1_{-0.3}^{+0.3} $  & $\lesssim0.7$             & ... & $ 0.95 (53 )$\\
Soft-state sample & CompTT+BB(cold) & $\lesssim0.5 $            & $2.5_{-0.1}^{+0.1} $ & $6.08_{-0.29}^{+0.25}$ & $1.24_{-0.04}^{+0.04}$ & $ 164.1_{-25.4}^{+27.9}$ & $ 1.07 (53)$\\
Soft-state sample & CompTT+BB(hot)  & $1.02_{-0.04}^{+0.03}$ & $2.6_{-0.1}^{+0.1}$  & $4.66_{-0.16}^{+0.16}$ & $\lesssim0.5$             & ... & $ 0.95 (53 )$\\
\enddata

\tablecomments{The detailed information of these two spectra are given
at the beginning of \S\ref{moddeg} and their unfolded spectra are
shown in Figures 4 and 5. The notations ``cold'' and ``hot'' refer to
the cold and hot-seed-photon models respectively (see
\S\ref{moddeg}). For CompTT+MCD models, a spherical geometry was
assumed for CompTT component, while for CompTT+BB models, a disk
geometry was assumed instead. $N_{\rm mcd}$ and $N_{\rm bb}$ are
normalizations (in unit of km$^2$ at distance 10 kpc) of MCD and BB,
respectively.}

\end{deluxetable}

\end{document}